\documentclass[lettersize,conference,10pt]{IEEEtran}
\IEEEoverridecommandlockouts
\usepackage{cite}
\usepackage{amsmath,amssymb,amsfonts,bbm}
\usepackage{bm}
\usepackage{nicefrac}
\usepackage{algorithmic}
\usepackage{textcomp}
\usepackage{xcolor}
\usepackage{hyperref}
\usepackage[utf8]{inputenc} 
\usepackage[T1]{fontenc}
\usepackage{graphicx}
\usepackage[font=small]{caption}
\usepackage{tikz}
\usetikzlibrary{positioning}
\usepackage{pgfplots}
\usepackage{epstopdf}
\definecolor{darkgreen}{rgb}{0.0, 0.5, 0.0}
\usepackage[labelformat=simple]{subcaption}
\usepackage{multirow}
\usepackage{lipsum}
\usepackage{mathtools}

\def\BibTeX{{\rm B\kern-.05em{\sc i\kern-.025em b}\kern-.08em
    T\kern-.1667em\lower.7ex\hbox{E}\kern-.125emX}}

\input{mysymbol.sty}

\pgfplotsset{compat=1.18}

\title{Ant Backpressure Routing for Wireless Multi-hop Networks with Mixed Traffic Patterns
}

\author{\IEEEauthorblockN{Negar Erfaniantaghvayi$^{*\dag}$, Zhongyuan Zhao$^{*\dag}$, Kevin Chan$^\ddag$,  Gunjan Verma$^\ddag$, \\
Ananthram Swami$^\ddag$, 
and Santiago Segarra$^\dag$
\thanks{Research was sponsored by the Army Research Office and was accomplished under Cooperative Agreement Number W911NF-24-2-0008 and W911NF-19-2-0269. 
The views and conclusions contained in this document are those of the authors and should not be interpreted as representing the official policies, either expressed or implied, of the Army Research Office or the U.S. Government. 
The U.S. Government is authorized to reproduce and distribute reprints for Government purposes notwithstanding any copyright notation herein.
\newline
$^*$ These authors contributed equally to this work.}
\thanks{Emails: $^\dag$\{ne12, zhongyuan.zhao, segarra\}@rice.edu, $^\ddag$\{kevin.s.chan.civ, gunjan.verma.civ, ananthram.swami.civ\}@army.mil}
\IEEEauthorblockA{$^\dag$\textit{Rice University}, Houston, TX, USA} 
\IEEEauthorblockA{$^\ddag$\textit{DEVCOM Army Research Laboratory}, Adelphi, MD, USA}
}}

\begin{document}

\maketitle
\pagestyle{plain} 

\begin{abstract} 
A mixture of streaming and short-lived traffic presents a common yet challenging scenario for Backpressure routing in wireless multi-hop networks.
Although state-of-the-art shortest-path biased backpressure (SP-BP) can significantly improve the latency of backpressure routing while retaining throughput optimality, it still suffers from the last-packet problem due to its inherent per-commodity queue structure and link capacity assignment. 
To address this challenge, we propose Ant Backpressure (Ant-BP), a fully distributed routing scheme that incorporates the multi-path routing capability of SP-BP into ant colony optimization (ACO) routing, which allows packets of different commodities to share link capacity in a first-in-first-out (FIFO) manner.
Numerical evaluations show that Ant-BP can improve the latency and delivery ratio over SP-BP and ACO routing schemes, while achieving the same throughput of SP-BP under low-to-medium traffic loads.
\end{abstract}


\section{Introduction}
Wireless multi-hop networks are widely used in military communications, disaster relief, and wireless sensor networks, and are envisioned to support emerging applications such as connected vehicles, robotic swarms, xG (device-to-device, wireless backhaul, integrated access and backhaul, and space-air-ground networks), Internet of Things, and machine-to-machine communications \cite{sarkar2013ad,kott2016internet,Cudak2021,akyildiz20206g}.
A key feature of these networks is their self-organizing capability enabled by distributed resource allocation schemes, without relying on centralized infrastructure.
Among these schemes, backpressure (BP) routing is a family of multi-path routing schemes that operate across physical, media access control (MAC), and network layers \cite{tassiulas1990stability,georgiadis2006resource,neely2005dynamic,zhao2023icassp, zhao2023enhanced, zhao2024bp,jiao2015virtual, ji2012delay,liaskos2023analysis}. 
In BP routing, each node maintains separate queues for packets of different destinations (also referred to as commodities). The routing decision selects the optimal commodity with the largest differential queue backlog (backpressure) across each link, and a set of non-interfering links with maximum total utility is scheduled for transmission (Max-Weight scheduling \cite{tassiulas1990stability,joo2011local}), where the link capacities are wholly assigned to the optimal commodities. 
This congestion-driven approach allows packets to explore various routes towards their destinations, achieving throughput optimality. 
However, due to slow startup, random walks, and the last-packet problem, BP routing suffers from poor latency, especially under low-to-medium traffic loads~\cite{jiao2015virtual,ji2012delay}. 
Although the first two drawbacks have been mostly mitigated by recent advances in shortest path-biased BP (SP-BP) routing  \cite{neely2005dynamic, georgiadis2006resource,jiao2015virtual, zhao2023icassp, zhao2023enhanced, zhao2024bp}, the last-packet problem persists due to commodity-dedicated route and link capacity assignments inherent to BP routing.

Real-world traffic in wireless multi-hop networks often comprises a mixture of streaming and bursty flows; the latter could be starved under throughput-oriented routing and scheduling schemes~\cite{jiao2015virtual,ji2012delay}. 
To address this challenge, we seek the best of both worlds of BP routing~\cite{tassiulas1990stability,georgiadis2006resource,neely2005dynamic,zhao2023icassp, zhao2023enhanced, zhao2024bp,jiao2015virtual,ji2012delay,liaskos2023analysis} and ACO routing~\cite{di2005anthocnet, zhang2017survey,dorigo2019ant} by proposing Ant Backpressure (Ant-BP) routing.
Like ACO routing, Ant-BP allows different commodities to share link capacity in a FIFO manner, while leveraging virtual SP-BP routing for route establishment.
Specifically, a newly arriving packet on a node is probabilistically assigned to one of the outgoing links based on their per-commodity pheromone values, established as the total number of packets across each link for each commodity during  virtual SP-BP routing.
Compared with sending out scout ants in conventional ACO routing, 
our virtual SP-BP routing has significantly lower overhead since neighboring nodes only exchange the number of virtual packets for each commodity rather than the actual payload in the stage of data transmission of virtual SP-BP.
By contrast, in ACO routing such as AntHocNet~\cite{di2005anthocnet}, a scout ant needs to record its forward path and travel backward from its destination for link pheromone updating, leading to high overhead.
Moreover, by modifying the statistical properties of virtual traffic, our approach can improve the quality of route establishment for short-lived traffic under the interference of streaming traffic.

\noindent
\textbf{Contribution.} The contributions of this paper are threefold:
\begin{enumerate}
    \item We propose Ant-BP, a novel routing scheme that combines the advantages of shortest path, backpressure, and ACO routing schemes, to mitigate resource starvation of short-lived flows under mixed traffic.
    \item We study the impact of design choices of virtual traffic during route establishment on both streaming and bursty flows under mixed traffic settings.
    \item Our numerical experiments demonstrate the effectiveness and robustness of our approach in improving packet delivery ratio and end-to-end latency for mixed traffic over both SP-BP and ant-based routing schemes.
\end{enumerate}
{\bf Notation.} 
In this paper, 
$ (\cdot)^\top $, $ \odot $, and $ |\cdot| $ represent the transpose operator, Hadamard (element-wise) product operator, and the cardinality of a set.
$ \mathbbm{1}(\cdot) $ is the indicator function. 

\section{System model}\label{sec:system}
We use an undirected connectivity graph $\mathcal{G}^n = (\mathcal{V}, \mathcal{E})$ to represent a wireless multi-hop network, in which $\mathcal{V}$ denotes a collection of nodes as user devices in the network, and $\mathcal{E}$ represents a set of links.
Each link $e = (i, j) \in \mathcal{E}$, where $i, j \in \mathcal{V}$, indicates that nodes $i$ and $j$ can directly communicate with each other. 
We assume $\mathcal{G}^n$ to be connected, ensuring that any two nodes within the network can reach each other.
To describe routing, we denote a directed link as  $(\overrightarrow{ i, j})$, representing data packets being transmitted from node $i$ to node $j$ via wireless link $(i, j)\in\ccalE$.
Moreover, we consider a set of flows $\mathcal{F}$ within the network. 
A flow $f = (i, c) \in \mathcal{F}$, where $i \neq c$ and $i, c \in \mathcal{V}$, represents data traffic originating from node $i$ and destined for node $c$, possibly through multiple hops. 
The packets destined for node $c$ are designated as commodity $c$.

We introduce the conflict graph, $\mathcal{G}^c = (\mathcal{E}, \mathcal{C})$, to depict the conflict relationships among links, defined as follows: 
each vertex $e \in \mathcal{E}$ corresponds to a wireless link within the network, and the presence of an undirected edge $(e_1, e_2) \in \mathcal{C}$ indicates the existence of an interference relationship between links $e_1$ and $e_2$.
We primarily focus on the interface conflict model, where the conflict graph is derived from the line graph of the connectivity graph. 
This represents the case where two links sharing the same node cannot be turned on simultaneously, e.g., if each node is equipped with only one radio transceiver.
However, our approach also applies to other conflict models, such as the physical distance interference model.
For the rest of this paper, we assume the conflict graph $\mathcal{G}^c$ to be known, either through direct monitoring of the wireless channel by each node~\cite{zhao2022link} or more advanced estimation~\cite{yang2016learning}.

The MAC of the wireless multi-hop network is considered to be a synchronized, time-slotted system with orthogonal multiple access. 
A time slot comprises a stage of route assignment and scheduling, followed by a stage of data transmission.
The total number of time slots is denoted by $T$.
The non-negative integer matrix $\mathbf{A} \in \mathbb{Z}_+^{|\mathcal{F}|\times T}$ collects the exogenous packet arrivals in real-time, where $\mathbf{A}_{f,t}$ represents the number of packets arriving at the source node of flow $f$ at time slot $t$. 
The matrix $\mathbf{R} \in \mathbb{Z}_+^{|\mathcal{E}|\times T}$ gathers the stochastic real-time link rates. Each entry $\mathbf{R}_{e,t}$ denotes the number of packets that can traverse link $e\in\ccalE$ during time slot $t$ and is presumed to exhibit symmetry in both directions.
The long-term link rates $\bbr\in\reals^{|\ccalE|}$, where $\bbr_e=\lim_{T\rightarrow\infty}\mathbb{E}_{t\leq T}(\bbR_{e,t}) $ for link $e\in\ccalE$.

\subsection{Per-Neighbor Queueing System}\label{sec:Queuesystem}

In our proposed routing scheme, each device $i\in\ccalV$ hosts per-neighbor queues, denoted as $\ccalQ_{ij}$ for all $j\in\ccalN_{\ccalG^n}(i)$, for packets to be sent to each neighboring device $j$ through link $(\overrightarrow{ i, j})$, where $ \ccalN_{\ccalG}(i) $ represents the set of immediate neighbors of node $i$ on graph $\ccalG$.
In addition, we use queue $\ccalQ_{ii}$ to buffer newly arrived packets from user $i$ (exogenous) or other devices (endogenous) with the next hops yet to be decided.
All the queues follow the FIFO principle, and the queue length of $\ccalQ_{ij}$ at time step $t$ is denoted by $q_{ij}(t)$.

\subsection{Per-Commodity Virtual Queueing System}\label{sec:vqueue}
Besides the physical queueing system, we also adopt a per-commodity virtual queueing system for virtual routing.
This entails a set of virtual queues denoted as $\{\tilde{\ccalQ}_i^{(c)} | i, c \in \mathcal{V}\}$, where $ \tilde{\ccalQ}_i^{(c)} $ is the virtual queue hosted on device $i$ for commodity $c$.
A virtual queue only records the number of virtual packets for the corresponding commodity without actually storing any packets, and the length of the virtual queue $ \tilde{\ccalQ}_i^{(c)} $  at virtual time step $ \tau $ is denoted as $\tilde{q}_i^{(c)}(\tau)$. 
Here, $\tau$ is adopted for virtual routing as it operates on a different time scale.

\begin{figure*}[!t]
    \includegraphics[width=0.95\linewidth]{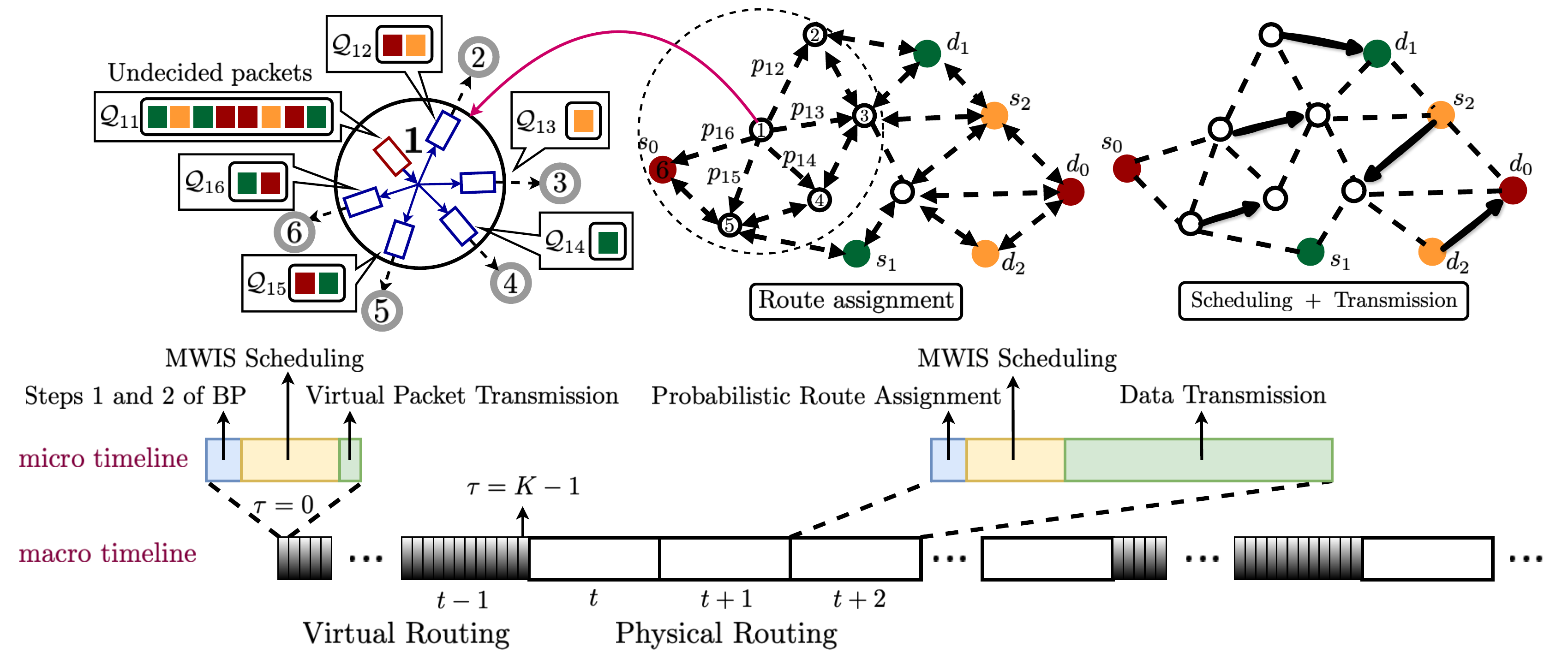}
    \vspace{-0.1in}
    \caption{The system diagram of Ant Backpressure routing: queueing system design, operations, and timelines.
    }
    \label{fig:timeline}
    \vspace{-0.1in}
\end{figure*}



\section{Ant Backpressure}
To mitigate the drawbacks of BP routing in the last-packet problem and lack of link capacity sharing among different commodities, our Ant-BP scheme adopts the per-neighbor queueing system described in Section~\ref{sec:Queuesystem}.

Recall that we use queue $\ccalQ_{ii}$ to buffer newly arrived exogenous and endogenous packets of which the next hops are still undecided.
The whole process is illustrated in the micro timeline in Fig.~\ref{fig:timeline} as ``Physical routing" and explained as follows.
In the beginning of a time slot $t$, each packet from $\ccalQ_{ii}$ on device $i\in\ccalV$ is sent independently to queue $\ccalQ_{ij}$ associated to one of its neighbors $j\in\ccalN_{\ccalG^n}(i)$ according to an established per-commodity routing policy, which is a probability distribution  defined on all outgoing links~\cite{zhang2017survey}
\begin{equation}\label{eq:policy}
p^{(c)}_{ij}(t) = \frac{\rho^{(c)}_{ij}(t)}{\sum_{l\in\ccalN_{\ccalG^n}(i)} \rho^{(c)}_{il}(t)}\;,    
\end{equation}
where $\rho^{(c)}_{ij}(t)$ is the pheromone value for commodity $c$ on link $(\overrightarrow{i,j})$ at time step $t$.
Notice that $\rho^{(c)}_{ij}(t)$ is updated via periodical virtual routing explained in Sections~\ref{sec:BP}and~\ref{sec:vrouting}, and remains the same across time steps until its next update. 

Following route assignment, the utility of each bidirectional link $(i,j)$ is calculated as~\cite{joo2011local,zhao2022link,zhao2023enhanced,zhao2023icassp,zhao2024bp}
\begin{align}\label{eq:utility}
    \bbu_{ij}(t)=\max\{q_{ij}(t),q_{ji}(t)\}\mathbf{R}_{ij,t}\;.
\end{align}
The direction of transmission for link $(i,j)$ is selected accordingly by the $\max$ operation.
We then run Max-Weight scheduling to find the schedule based on the utility vector $\bbu(t)\in\reals_{+}^{|\ccalE|}$ and the conflict graph $\ccalG^c=(\ccalE,\ccalC)$, 
\begin{subequations}\label{eq:mwis}
\begin{align}
    & \bbs^*(t) = \argmax_{\bbs(t) \in \{0, 1\}^{|\mathcal{E}|}} \bbs(t)^\top \bbu(t),\\
    \text{s.t. }& \bbs_{e_1}(t) + \bbs_{e_2}(t)\leq 1,\; \forall\; (e_1, e_2)\in \ccalC \;.
\end{align}
\end{subequations}
Notice that the Max-Weight scheduling in (\ref{eq:mwis}) involves solving a maximum weighted independent set (MWIS) problem on the conflict graph, which is NP-hard \cite{joo2010complexity}. 
In this paper, we choose local greedy scheduler (LGS)~\cite{joo2011local} as our distributed Max-Weight scheduler for its simplicity. 
Based on the link utility defined in \eqref{eq:utility}, the number of packets transmitted on an activated link $(\overrightarrow{ i, j})$ is $\min\{q_{ij}(t), \mathbf{R}_{ij,t}\}$.
Next, we explain the establishment of a routing policy.

\subsection{Virtual SP-BP Routing for Pheromone Establishment}
\label{sec:BP}

The virtual routing adopts shortest path-biased backpressure (SP-BP) routing~\cite{neely2005dynamic,georgiadis2006resource,zhao2023icassp,zhao2023enhanced,zhao2024bp}, based on the per-commodity queueing system described in Section~\ref{sec:vqueue}. 
In each time step $\tau$, the virtual SP-BP comprises the following four steps: 
First, the optimal commodity $c^*_{ij}(\tau)$ for each directed link $(\overrightarrow{ i, j})$ is selected as the one with the maximal backpressure:
\begin{equation}\label{eq:optimalcomm}
    c^*_{ij}(\tau) = \underset{c \in V}{\text{argmax}} \{ \eta^{(c)}_i(\tau) - \eta^{(c)}_j(\tau) \},
\end{equation}
where $\eta^{(c)}_i(\tau) = \tilde{q}^{(c)}_i(\tau) + B^{(c)}_i$, $\tilde{q}^{(c)}$ is the virtual queue length, and $B^{(c)}_i \geq 0$ is a queue-agnostic shortest path bias for each pair of nodes and commodity $i, c \in \mathcal{V}$. 
In this paper, the set of biases $\{B^{(c)}_i | i,c\in\ccalV\} $ is established through all-pairs-shortest-path (APSP) algorithm on the connectivity graph $\ccalG^n$ with edge weights $\left[ \delta_e |e\in\ccalE \right]$ defined by link features~\cite{zhao2023icassp,zhao2023enhanced,zhao2024bp}, e.g., $ \delta_e=\bar{r}r_{\max}/\bbr_e $, where  $\bar{r}=\mathbb{E}_{e\in\ccalE}(\bbr_e)$, and $ r_{\max}=\max_{e\in\ccalE} \bbr_e $.
To accommodate network mobility, $\{B^{(c)}_i | i,c\in\ccalV\} $ can be updated periodically.
In step 2, the maximum differential backlog of $(\overrightarrow{ i, j})$ is found as
\begin{align}
\label{eq:backlog}
    \bbw_{ij}(\tau) \!=\! \max\!\left\{ \eta^{(c^*_{ij}(\tau))}_i\!(\tau) \!-\! \eta^{(c^*_{ij}(\tau))}_j\!(\tau),\!0\!\right\}.
\end{align}
In step 3, the schedule $\bbs(\tau) \in \{0, 1\}^{|\mathcal{E}|}$ is found as the set of non-conflicting links with the maximum total utility, 
\begin{subequations}
\begin{align}
    & \bbs^*(\tau) = \argmax_{\bbs(\tau) \in \{0, 1\}^{|\mathcal{E}|}} \bbs(\tau)^\top [\mathbf{R}_{*,\tau} \odot \widetilde{\bbw}(\tau)],  \label{eq:bp:mwis}\\
    \text{s.t. }& \bbs_{e_1}(\tau) + \bbs_{e_2}(\tau)\leq 1,\; \forall\; (e_1, e_2)\in \ccalC \;.
\end{align}    
\end{subequations}
where vector $\bbR_{*,\tau}=[\bbR_{e,\tau}|e\in\ccalE]$, the virtual real-time link rate of link $e$ at step $\tau$, $\bbR_{e,\tau}$, follows the same distribution of $\bbR_{e,t}$ to reflect the characteristics of underlying wireless channel; 
vector $\widetilde{\bbw}(\tau) = [\widetilde{\bbw}_{ij}(\tau)|(i, j) \in \mathcal{E}]$ with $\widetilde{\bbw}_{ij} = \max\{\bbw_{ij}(\tau), \bbw_{ji}(\tau)\}{\cdot\mathbbm{1}\!\!\left[\!\tilde{q}_i^{(c_{ij}^*(\tau))}\!(\tau)\!>\!0\!\right]}$~\cite{zhao2024bp}, and the direction of the link selected by the max function will be recorded for step 4; the per-link utility in Max-Weight scheduling is $\bbu_{ij}(\tau) = \mathbf{R}_{ij,t} \widetilde{\bbw}_{ij}(\tau)$. 
The Max-Weight scheduler is also selected as LGS in~\cite{joo2011local}.
In step 4, all of the virtual real-time
link rate $\mathbf{R}_{ij,\tau}$ of a scheduled link is allocated to its optimal commodity $c^*_{ij}(\tau)$. 
The final routing and transmission assignments of commodity $c \in \mathcal{V}$ on link $(\overrightarrow{ i, j})$ is
\begin{align}
\label{eq:trans}
    \mu^{(c)}_{ij}(\tau) = \begin{cases} \mathbf{R}_{ij,\tau}, & \text{if } c = c^*_{ij}(\tau), \bbw_{ij}(\tau) > 0, s_{ij}(\tau) = 1, \\ 0, & \text{otherwise.} \end{cases}
\end{align}

The number of virtual packets of commodity $c$ travelling across link $(\overrightarrow{ i, j})$ accumulates as follows: $ n_{ij}^{(c)}(0)=0 $, and $$ n_{ij}^{(c)}(\tau) = (1-\varepsilon)\cdot n_{ij}^{(c)}(\tau-1) + \min\{\tilde{q}^{(c)}_{i}(\tau), \mu^{(c)}_{ij}(\tau)\}, $$ where the evaporation rate is typically set as zero ($\varepsilon=0$).
The pheromone intensity in \eqref{eq:policy} is established as 
\begin{equation}\label{eq:pheromone}
    \rho^{(c)}_{ij} = {\max\{n^{(c)}_{ij}(K) - n^{(c)}_{ji}(K), 0\}} + \epsilon\;,
\end{equation} 
where $n^{(c)}_{ij}(K)$ is the total number of virtual packets of commodity $c$ travelled across link $(\overrightarrow{ i, j})$ during virtual SP-BP routing, and 
$K$ is the required number of time steps for virtual SP-BP routing to establish our routing policy, which is configured as a system parameter via trial-and-error. 
The small positive constant $\epsilon>0$ in \eqref{eq:pheromone} ensures $ p^{(c)}_{ij}(t) = 1/|\ccalN_{\ccalG^n}(i)|$ when the first term in \eqref{eq:pheromone} is zero for all $ l\in\ccalN_{\ccalG^n}(i) $.

In virtual SP-BP routing, only the number of virtual packets of each commodity stored at each node $i\in\ccalV$ and transmitted over each link $(\overrightarrow{i,j})$ is tracked and exchanged across the network, no packets are actually generated or transmitted. 
As a result, the data transmission of virtual routing can be compressed into a very short duration, such that tens or hundreds of virtual time steps can be finished in a single physical time slot.
This allows our routing policy to be established within a few time slots, as illustrated by the timelines in Fig.~\ref{fig:timeline}.

\subsection{Virtual Traffic Configuration}\label{sec:vrouting}

In virtual routing, we generate a virtual flow for each physical flow with the same source and destination, but potentially different configurations of flow rate $\lambda$ and duration. 
For example, a bursty physical flow lasting only tens of time slots can be treated as a streaming flow with a different flow rate.
Based on the knowledge of the flow rate and duration (streaming vs. bursty) of physical flows, we can either copy the exact flow rate and duration of the physical flow or configure all virtual flows as streaming type with moderate flow rate.
Our experiment in Section~\ref{sec:results} shows that setting all virtual flows as streaming type can yield a better routing policy, and the mismatch in flow rate between virtual and physical flows has minimal impact on the performance. 

Even if a virtual flow is configured exactly the same as its physical flow, it is impossible to replicate the exact arrival process of each physical packet since virtual traffic occurs before the physical traffic.
Therefore, the arrival process of virtual traffic at its source node follows a Poisson process with the configured arrival rate (flow rate).

\section{Numerical Experiments}\label{sec:results}
To assess the performance and robustness of our proposed Ant-BP, we conducted evaluations in simulated wireless multi-hop networks. These evaluations included comparisons with several benchmarks, including the state-of-the-art SP-BP~\cite{zhao2024bp} and two ACO routing schemes.

\subsection{Test Setup and Benchmarks}\label{sec:results:config}
Our simulated wireless networks are created from a 2D point process, generating $|\ccalV|=100$ nodes 
randomly spread across a squared area with a uniform density of $8/\pi$.
A wireless link between two nodes is formed if they are positioned within a distance of $1$ unit, and the conflict graph is established by interface conflict model, representing a simplified scenario that all devices use  uniform transmit power with beamforming.
In this test configuration, the average degree of conflict graphs is $15.4$.
For each network size $|\ccalV|$, we generate $100$ test instances by drawing $10$ instances of random networks, each with $10$ realizations of random source-destination pairs and random link rates. 
Each test instance encompasses a number (uniformly chosen between $\lfloor 0.30 |\ccalV|\rfloor$ and $ \lceil 0.50|\ccalV|\rceil$) of random flows between various source-destination pairs.
To capture fading channels with lognormal shadowing, the long-term link rate of a link is uniformly distributed as $\bbr_{e} \sim \mathbb{U}(10,42)$, and the real-time link rates, $\bbR_{e,t} \sim  \mathbb{N}(\bbr_{e}, 3)$, truncated within a range of $\bbr_e\pm 9$.
A test instance is simulated over a duration of $T=1000$ time steps.

We define two traffic types: streaming and bursty flows.
For a streaming flow, the exogenous packet arrival rate follows a Poisson process with a rate of $L_{s}\lambda_{s} $, where $\lambda_{s}\in \mathbb{U}(0.2, 1.0)$ throughout the simulation.
For a bursty flow, the exogenous packet arrival rate follows a Poisson process with a rate of $L_b\lambda_b$, where $\lambda_{b}\in \mathbb{U}(0.2, 1.0)$ for $t<30$ and $\lambda_b=0$ for $t\geq 30$. 
Here, $L_{b}$ and $L_s$ denote the bursty and streaming traffic loads, respectively.
Under mixed traffic setting, a flow is configured as bursty with a probability of $P_{b}=0.5$.

\begin{figure*}[t!]
    \hspace{-3mm}
    \subfloat[]{
    \includegraphics[width=0.33\linewidth]{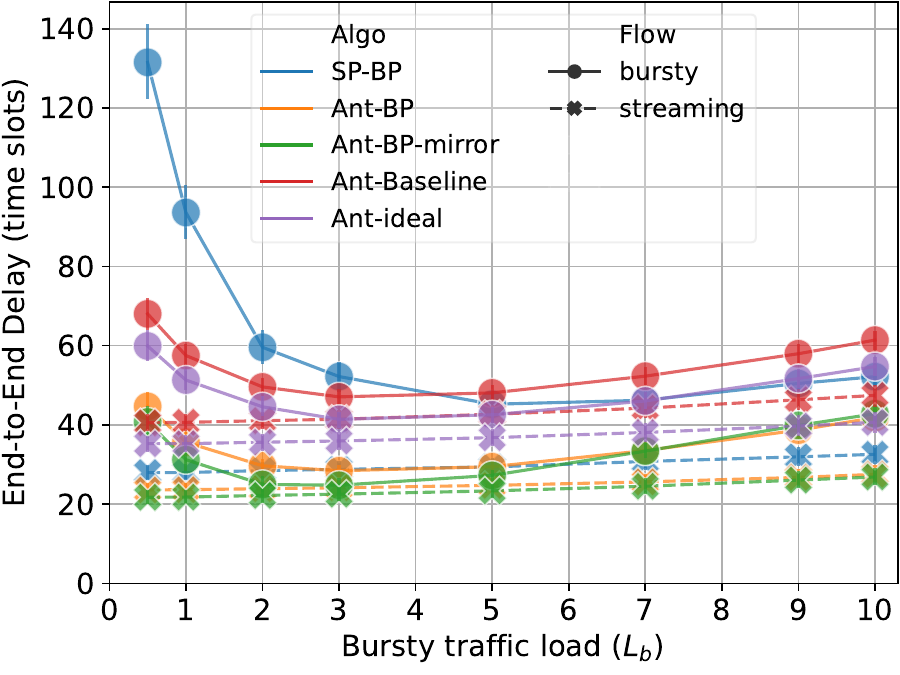}
    \label{fig:mixed:delay}
    }
    \hspace{-3mm}
    \subfloat[]{
    \includegraphics[width=0.33\linewidth]{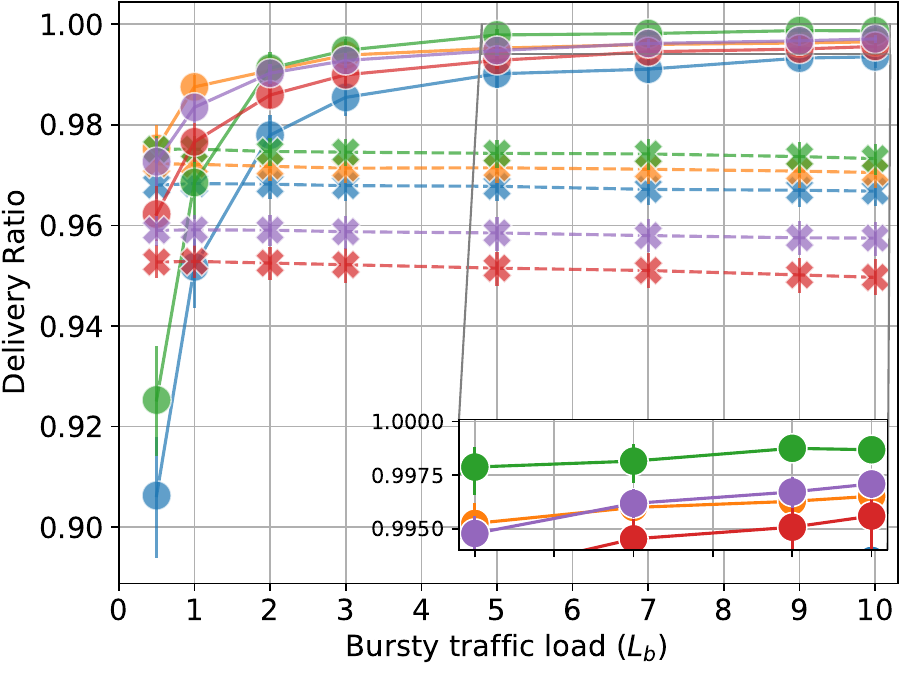}
    \label{fig:mixed:delivery}
    }
    \hspace{-3mm}
    \subfloat[]{
    \includegraphics[width=0.33\linewidth]{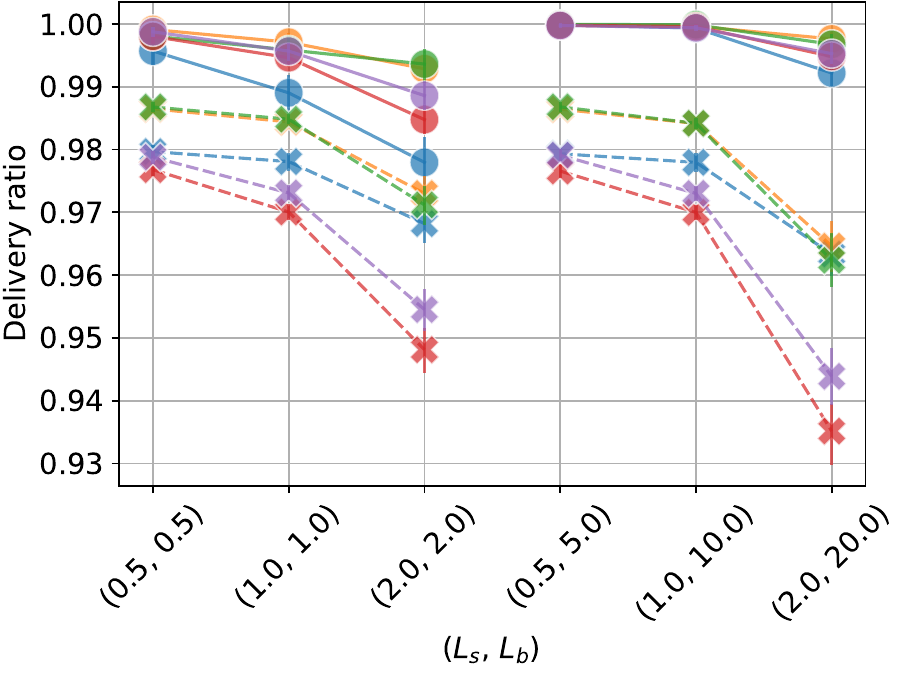}\label{fig:robust}\label{fig:robust:delivery}
    }  
    \vspace{-0.1in}
    \caption{The average (a) end-to-end latency and (b) delivery ratio of tested routing policies as a function of bursty flow load $ L_b $ in 10 random instances of wireless networks of 100 nodes under mixed traffic setting with a constant streaming load  $L_s=2.0$. 
    (c) The average delivery ratio of tested routing schemes in wireless networks of 100 nodes under mixed traffic with different loads shown in the x-axis. 
    For Ant-based schemes, routing policies are established based on virtual traffic flows of $(1.0, 1.0)$ and $(1.0, 10.0)$ for the left and right sides, respectively.} 
    \label{fig:mixed}
    \vspace{-0.1in}
\end{figure*}

We evaluate two variations of Ant-BP along with three other benchmarks:
1) \textit{Ant-BP}: all virtual flows are configured as streaming flows with traffic load $L_s$, regardless of the traffic type and flow rates of their corresponding physical flows.
2) \textit{Ant-BP-mirror}: a virtual flow has the exact flow rate and traffic type of its corresponding physical flow.
For 1) and 2), the routing policy is established using SP-BP as described in~\cite{zhao2023enhanced,zhao2024bp}, where the shortest path biases $\{B_{i}^{(c)}|i,c\in\ccalV\}$ are established by APSP on $\ccalG^n$ with edge weights $ \delta_e=\bar{r}r_{\max}/\bbr_e, \forall e\in\ccalE$ as explained in Section~\ref{sec:BP}.
3) \textit{SP-BP}: the same SP-BP for 1) and 2) directly running on physical traffic.
4) \textit{Ant-baseline}: the same physical routing as Ant-BP, with pheromones established by virtual ACO routing and frozen during physical routing. 
For a fair comparison with Ant-BP, the routing probability in the virtual ACO routing phase is
\begin{align}
    \label{eq:sgs:test}
    p_{ij}^{(c)}(t) = \frac{\rho_{ij}^{(c)}(t) + h_{ij}^{(c)}}{\sum_{l\in\ccalN_{\ccalG^n}(i)} \left[\rho^{(c)}_{il}(t) + h_{il}\right]}\;,
\end{align}
where $h_{ij}^{(c)}=B_{i}^{(c)}-B_j^{(c)}$ is the heuristic link cost based on the same $\{B_i^{(c)}\}$ used for Ant-BP, and time step $t$ indicates that virtual ACO routing operates on the same time scale of physical routing. 
The pheromone in~\eqref{eq:sgs:test} is updated as~\cite{zhang2017survey}
\begin{align}\label{eq:ph_update}
    \rho^{(c)}_{ij}(t+1) &= (1 - \varepsilon) \cdot \rho^{(c)}_{ij}(t) + \sum_{k=1}^{m(c,t)} \theta_{ij,k}^{(c)}(t),
\end{align}    
where $\varepsilon=0.002$ is the evaporation rate, $m(c,t)$ is the number of ants arriving destination $c$ at time step $t$, and
$\theta_{ij,k}^{(c)}(t)$ is the amount of pheromone deposited by the $k$th ant.
For Ant-baseline, $ \theta_{ij,k}^{(c)}(t) =0.01$ is a constant, therefore pheromone update in \eqref{eq:ph_update} occurs at the end of time step $t$, with $m(c,t)$ as the number of ants of commodity $c$ across $(\overrightarrow{i,j})$ in step $t$. 
5) \textit{Ant-Ideal}: the same physical routing as Ant-BP, with pheromone established by virtual ACO routing {using \eqref{eq:sgs:test}} and maintained by proactive ants in physical routing.
During physical routing, a proactive ant is sent out for every 100 data packets, and has a $10\%$ chance to explore the routes by uniformly selecting an outgoing link.
For Ant-Ideal, the pheromone update in \eqref{eq:ph_update} occurs instantaneously at the end of each virtual or physical time step without ants traveling backward to inform the nodes on their paths.
The pheromone deposit of ant $k$ in~\eqref{eq:ph_update} for Ant-Ideal is~\cite{zhang2017survey}
\begin{align}  \label{eq:ph_update:delta}  
    \theta_{ij,k}^{(c)}(t) &= \begin{cases} 
    \left[\phi\big(\ccalP^{(c)}_k\big)\right]^{-1}, & \text{if } (\overrightarrow{ i, j})\in \ccalP^{(c)}_k \\ 
    0, & \text{if } (\overrightarrow{ i, j})\notin \ccalP^{(c)}_k  
    \end{cases},
\end{align}   
where $\ccalP^{(c)}_k $ is the path of the $k$th ant, and function $\phi\big(\ccalP^{(c)}_k\big)$ is the end-to-end latency of ant $k$.
For Ant-Baseline and Ant-Ideal, pheromone intensities on all links are initialized as $ \rho^{(c)}_{ij}(0) = 1.3$ in virtual routing.
The total number of time steps in virtual routing for all relevant schemes is $K=1000$.
Notice that even with the same $K$ steps of virtual routing, the physical overhead of Ant-BP will be significantly shorter than Ant-Baseline and Ant-Ideal because the former only needs to send out the count of packets while the latter two have to send out actual packets.



\subsection{Performance under Mixed Traffic}\label{sec:results:mixed}
To evaluate the routing quality and effectiveness in handling short-lived traffic in the presence of streaming flows, we tested the mentioned five schemes under a mixed traffic setting with a fixed streaming load $L_s = 2.0$ and varying bursty loads $L_b \in \{0.5, 1, \dots, 10\}$.
The average end-to-end latency and delivery ratio as a function of $L_b$ are shown in Figs.~\ref{fig:mixed:delay} and~\ref{fig:mixed:delivery}, respectively.

Given a constant $L_s$, as $L_b$ increases, the total number of packets injected into the networks increases slightly (i.e., by a factor of $L_b/33.3$), 
therefore, the average latency (delivery ratio) of streaming flows for all routing schemes only increases (decreases)  slightly, as illustrated by dashed lines in Figs.~\ref{fig:mixed:delay} and~\ref{fig:mixed:delivery}.
Ant-BP-mirror and Ant-BP rank the best and second best in terms of latency and delivery ratio for streaming flows, followed by SP-BP, Ant-ideal, and Ant-Baseline. 
Their average delivery ratios are $0.974, 0.971, 0.968, 0.958$, and $0.952$, respectively. 
As illustrated by the solid lines in Fig.~\ref{fig:mixed:delay}, the latency of bursty flows is always higher than that of streaming flows, and more sensitive to $L_b$,  
since bursty traffic lacks consistent congestion gradients to drive the scheduling of links on its routes in the presence of streaming flows under queue length-based Max-Weight scheduling.
However, since all packets of bursty flows arrive in the first $30$ time slots of the simulation, most of them reach their destinations after spending sufficient time in the network $T=1000$, resulting in a higher delivery ratio than streaming traffic, as illustrated in Fig.~\ref{fig:mixed:delivery}.
The fact that Ant-BP and Ant-BP-mirror almost always outperform Ant-Baseline and Ant-ideal in latency and delivery ratio for streaming and bursty flows shows that the pathfinding based on virtual SP-BP is superior to ACO. 
The gain of Ant-BP and Ant-BP-mirror over SP-BP also demonstrates the effectiveness of FIFO-based link capacity sharing among commodities.

For lower burst loads $L_b \leq 3$, the latency and delivery ratio of bursty traffic for all routing schemes get worse as $L_b$ decreases, as illustrated in Figs.~\ref{fig:mixed:delay} and~\ref{fig:mixed:delivery}, presenting the most challenging cases where bursty traffic is starved from link scheduling. 
In particular, among all the schemes, SP-BP exhibits the worst latency (e.g., $131.5$ for $L_b=0.5$) and the lowest delivery ratio (e.g., $0.906$ for $L_b=0.5$) under bursty traffic conditions, due to its routing decisions also depending on congestion gradients.
Ant-BP achieves the best delivery ratio (e.g., $0.975$ for $L_b=0.5$) and ranks second in latency (e.g., $44.7$ for $L_b=0.5$) under bursty flows with $L_b \leq 3$. 
It does so by leveraging the superior pathfinding capability of virtual SP-BP and FIFO-based link capacity sharing.
Ant-BP-mirror ranks the best in delivery ratio and latency for $L_b\geq 2$ by further leveraging the exact knowledge of bursty flows. 
However, for $L_b=0.5$, its delivery ratio drops significantly to $0.925$ since its virtual SP-BP also suffers from the last-packet problem.
In contrast, Ant-BP can avoid this issue by configuring all the virtual flows as streaming, 
thus, it manages lightweight bursty traffic more effectively, at the cost of a slight degradation in latency and delivery ratio for streaming and heavier bursty traffic compared to Ant-BP-mirror.

Virtual routing may face the challenge of imperfect knowledge of physical traffic flow rates. 
To evaluate the robustness of routing policies learned from mismatched flow rates, we test four ant-based schemes under physical flow rates that are half or double of those used for route establishment ($L_s=1.0, L_b=1.0$ and $L_s=1.0, L_b=10.0$). 
For comparison, we also test SP-BP based on the exact queueing state information.
The overall trends of delivery ratio by traffic load and the rankings of tested schemes in Fig.~\ref{fig:robust} are consistent with the results in Fig.~\ref{fig:mixed:delivery}. 
Even with mismatched virtual flow rates, Ant-BP and Ant-BP-mirror generally maintain their leading ranks for both streaming and bursty traffic. 
In the special case of $L_s=2.0, L_b=20.0$, SP-BP achieves slightly lower latency than Ant-BP and Ant-BP-mirror for both streaming flows ($41.2$ v.s.  $47.6$ and $44.1$) and bursty flows ($79.8$ v.s. $92.3$ and $82.1$) due to its advantage of managing heavier traffic; 
nonetheless, its delivery ratios are similar to those of Ant-BP and Ant-BP-mirror in streaming traffic ($0.963$ v.s. $0.960$ and $0.962$) and slightly worse in bursty flows ($0.992$ v.s $0.995$ and $0.997$).
It shows that the routing policies of Ant-BP and Ant-BP-mirror are robust to mismatched virtual flow rate configurations.

\subsection{Throughput under Streaming Traffic}
\begin{figure}[!t]
    \centering
    \includegraphics[width=0.9\linewidth]{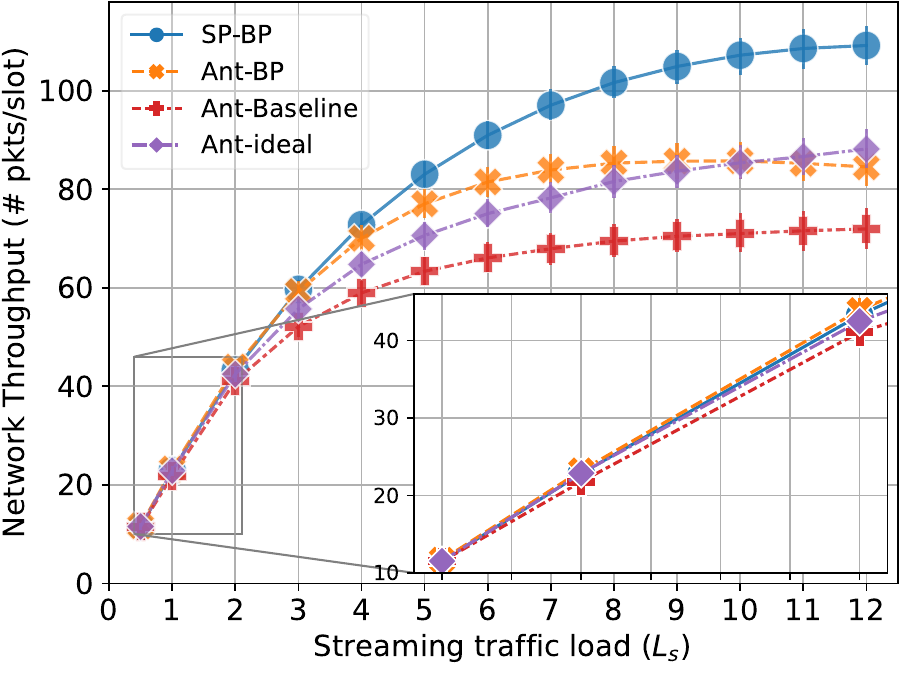}
    \caption{Average network throughput (network-wide number of packets delivered per time slot) of routing schemes on wireless networks with 100 nodes under all streaming traffic as a function of traffic load $L_s$.
    }
    \label{fig:throughput}
\end{figure}

To better understand the pros and cons of Ant-BP, we compare the throughput performance of the routing schemes under all streaming flows across different traffic loads $L_s \in \{0.5, 1, 2, \dots, 12\}$, by counting the number of delivered packets per physical time-slot across the network. 
In Fig.~\ref{fig:throughput}, Ant-BP achieves similar or slightly better throughput compared to SP-BP for lower traffic loads ($L_s\leq3$), but only $84.4\%$ of throughput of SP-BP for $L_s>3$ on average. 
This is because the link capacity sharing of Ant-BP brings limited benefits for streaming traffic, whereas the resource allocation decisions of SP-BP based on instantaneous queueing state information can better manage heavier traffic.
This result, along with observations from Fig.~\ref{fig:mixed}, demonstrates that Ant-BP can improve the performance of bursty traffic over SP-BP without compromising throughput under low-to-medium streaming traffic intensities ($L_s\leq 3$). 
Moreover, the throughput of Ant-ideal surpasses that of Ant-BP for higher traffic loads $L_s\geq 10$ due to its cost-based pheromone deposit mechanism and continuous route maintenance using proactive ants. 
It suggests that the throughput of Ant-BP might be further improved by incorporating such mechanisms of Ant-ideal.

\section{Conclusions and Future Work}
In this study, we propose to improve the performance of short-lived traffic in wireless multi-hop networks under the interference of streaming traffic, by incorporating the superior path-finding capability of SP-BP into ACO routing with FIFO-based link capacity sharing. 
Our approach is shown to be effective for mixed traffic, robust to mismatched virtual flow configuration, and can achieve similar throughput of SP-BP under low-to-medium traffic intensity.
However, the cost of these benefits is lower throughput under heavier traffic. 
Future directions include developing continuous route maintenance mechanisms to improve throughput under heavier traffic, and adapting to node mobility and link failure in wireless networks.

{\footnotesize
\bibliography{references}
}
\end{document}